\documentclass[
    reprint,
    superscriptaddress,
    amsmath,
    amssymb,
    aps,
]{revtex4-2}

\usepackage{graphicx}
\usepackage{float}
\usepackage{amsfonts,amsmath,bbm}
\usepackage[colorlinks=true,linkcolor=blue]{hyperref}
\usepackage{array}
\usepackage{comment}
\usepackage{xcolor}
\DeclareMathOperator{\sech}{sech}

\newcolumntype{P}[1]{>{\centering\arraybackslash}p{#1}}

\makeatletter
\def\frontmatter@abstractwidth{\textwidth}
\makeatother

\begin{document}

\title{Ferromagnetic Bloch Line as a Bead on a String}

\author{Jake Hoffmann}
\author{Oleg Tchernyshyov}

\affiliation{
William H. Miller III Department of Physics and Astronomy,
Johns Hopkins University,
Baltimore, Maryland 21218, USA
}

\date{\today}

\begin{abstract}

We present a simple model of a Bloch line in a thin-film ferromagnet with strong easy-axis and intermediate in-plane anisotropies. The meron is represented as a charged, massless particle that slides along the linear domain wall in a background magnetic field-- a gyroscopic bead on a string. Its gyroscopic behavior dynamically isolates the wall segments on either side of the meron, motivating the model for a Bloch line as two strings joined at a moving boundary. The meron clamps the wall at this junction, completely reflecting low-amplitude spin waves that propagate along it. Out-of-plane and in-plane magnetic fields and spin-transfer torques then act in two ways: by exerting a force on the Bloch line itself, or by driving the wall as a whole. Predictions of this model are in agreement with numerical simulations.

\end{abstract}

\maketitle

\section{Introduction}

In this paper we aim to understand the dynamics of a Bloch line in a ferromagnetic thin film. A Bloch line is a magnetic soliton living inside a domain wall, across which the in-plane magnetization of the wall rotates by $180^\circ$. Because this rotation carries the magnetization over half the unit sphere, the Bloch line is also a half-skyrmion, or meron, carrying topological charge $\pm \frac{1}{2}$.

Bloch lines have been of long-standing interest as information carriers in the development of high-density information storage devices \cite{ Humphrey:1985, S.S.P.Parkin:2008, Yang:2021}. The domain wall constrains the meron to move along a confined track, preventing drift, and the topological robustness of the meron allows it to achieve ultra-fast speeds \cite{Herranen:2015, Herranen:2017, Yang:2021}. Here, we develop a general description for how these merons interact with the wall that carries them, and of the dynamics that follow from it.

\begin{figure}[tbp]
    \centering
    \includegraphics[width=.9\columnwidth]{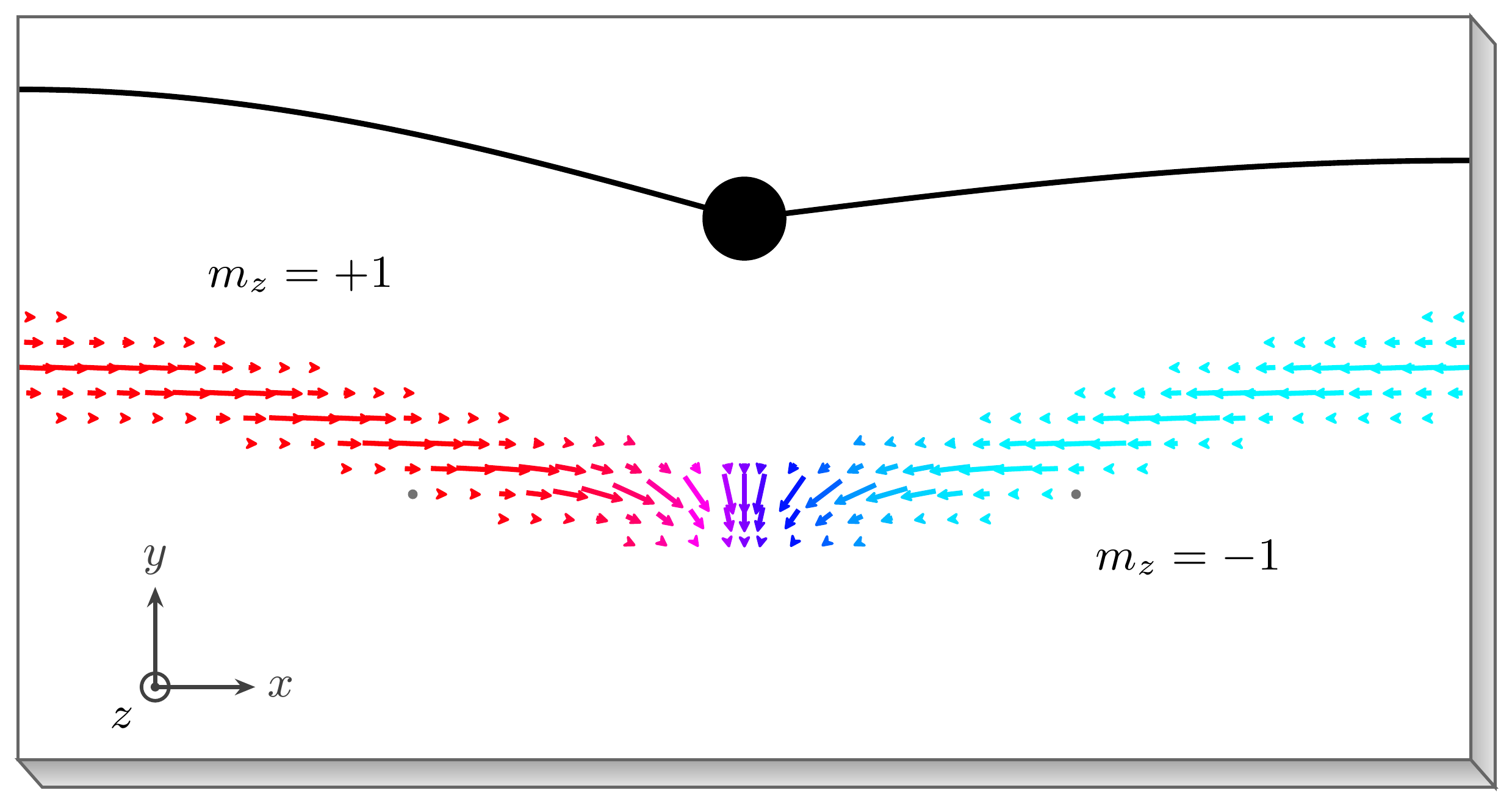}
    \caption{Representation of a meron on a domain wall as a bead on a string. The configuration illustrated here is the same one used for micromagnetic simulations, see Sec. \ref{sec:Mumax} for details.}
    \label{fig:BeadOnString}
\end{figure}

We expand on the earlier theoretical treatments of \cite{Malozemoff:1972, Slonczewski1974BlochLine, ZvezdinPopkov:1986, Nikiforov1986BlochLines} and use the method of collective coordinates to reduce the domain wall to a mechanical string-like object with tension $\sigma$ and inertia $\rho$ \cite{Zhang:2018}. The meron is represented as a charged, massless bead threaded along the domain wall, and all motion takes place in a background magnetic field \cite{Dasgupta:2020}. The domain wall segments to the left and right sides of the meron may have different tension and inertia, depending on the spin orientation. The possibility of differing parameters, combined with the gyroscopic dynamics of the meron, motivates us to model the system as two separate strings connected at a moving boundary rather than as a continuous string with a defect. This is illustrated in \textbf{Fig.} \ref{fig:BeadOnString}. 

Zvezdin and Popkov \cite{ZvezdinPopkov:1986} provide the closest precursor to this description. Like us, they treat the Bloch line as a point-particle and derive equations of motion containing the same gyroscopic meron-wall coupling and external-field driving terms that later appear here. The key difference is in how each treatment handles the domain wall away from the line. They impose a restoring force to keep the wall rigid away from the meron, so that spin waves emitted by the line are permanently lost and absorbed into an effective friction. Our approach instead retains the full wall dynamics so that disturbances propagate, reflect, and return to collide with the meron-- applying tension forces as they do so. 

It is this interplay between string tension and gyroscopic dynamics that governs the meron's motion. The two segments pull on the meron with their tension forces; unless these pulls balance, the meron experiences a net force, which the gyroscopic coupling converts from transverse push to longitudinal motion. This leads to one of the central results of this work: In the low-amplitude regime, the meron acts as a near-perfect clamp on the domain wall, preventing the transmission of spin waves, and dynamically isolating the segments on its left and right. 

Furthermore, the standard drives separate according to where the force is applied. An in-plane magnetic field or a current along the wall acts on the Bloch line itself, while an out-of-plane field or a current across the wall drives the domain wall as a whole. (An in-plane field also renders the two segments inequivalent, splitting their tensions and wave speeds.) This brings a common description to each perturbation and the resulting motion. Finally, every result is verified against micromagnetic simulations \cite{MuMax}. 

The paper is organized as follows. Section \ref{sec:BeadString} develops the bead-on-a-string model, in which the Bloch line is stabilized by strong out-of-plane and intermediate in-plane anisotropies and long-range dipolar interactions are omitted. We also derive the equations of motion and the clamping result here. Section \ref{sec:perturbations} explores the effects of several external perturbations, namely out-of-plane and in-plane magnetic fields and spin transfer torques. We find that in an out-of-plane magnetic field, the average spin orientation changes at the Larmor frequency, and an in-plane magnetic field will split the tension and inertia parameters between the left and right sides. In addition, we show that the meron may undergo extremely fast motion while remaining intact when exposed to a current along the domain wall, consistent with \cite{Herranen:2015, Herranen:2017, Yang:2021}. Section \ref{sec:Mumax} compares our theory with trajectories from MuMax3 simulations and finds close agreement. 

\section{Model as a Bead on a String}
\label{sec:BeadString}

\subsection{String Parameters}
\label{sec:string params}
The first order of business is to determine the tension and mass density associated with the bulk domain wall segments. The strategy follows \cite{Makhfudz:2012,Zhang:2018} closely: we parametrize the unit vector of magnetization $\mathbf{m} \equiv \mathbf{M}/M_s$ in terms of its polar and azimuthal angles $\theta$ and $\phi$ such that $\mathbf{m} = (\sin\theta \cos\phi, \, \sin\theta \sin\phi, \, \cos\theta)$. Tension is evaluated as the energy per unit length of a segment, $U = \sigma l$, and we integrate out $\phi$ from the domain wall Lagrangian to obtain the D{\"o}ring mass density $\rho$ \cite{Doring:1948}. 

Both parameters require a calculation of potential energy. At the first stage, two contributions enter.

\textit{Exchange:}
\begin{align}
U_{\mathrm{exchange}}
&= Ah \int d^2r \,
\partial_i\mathbf{m}\cdot\partial_i\mathbf{m}
\notag\\
&= Ah \int d^2r \,
\left[
(\nabla\theta)^2
+\sin^2\theta\,(\nabla\phi)^2
\right].
\label{eq:exchange}
\end{align}

\textit{Easy-axis anisotropy}, with the strong easy axis $\hat{z}$ normal to the plane of the film:
\begin{align}
U_{\mathrm{anisotropy,z}}
&= -K_z h \int d^2 r \, m_z^2
= K_z h \int d^2r \, \sin^2\theta.
\end{align}

The domain wall runs along the $x$ axis. Consider first a uniform wall, where the polar angle varies only across it, $\theta = \theta(y)$, and the azimuthal angle is constant at $\phi = \Phi$. The energy then reads

\begin{equation}
U = h \int dxdy \, \left[
A (\partial_y\theta)^2
+ K_z \sin^2\theta
\right].
\label{eq:TotalEnergy}
\end{equation}

Its minimization with respect to $\theta$ gives

\begin{align}
\cos \theta(y) = \zeta_w \tanh \frac{y - Y}{\lambda_{w}},
\quad \lambda_{w} = \sqrt{\tfrac{A}{K_z}},
\label{eq:Theta_eq}
\end{align}

where $\lambda_w$ is the width of the domain wall and $Y(x,t)$ is the wall position. The $\mathbb{Z}_2$ topological charge $\zeta_w \equiv \tfrac{1}{2} \int_{-\infty}^{\infty} dy\, \frac{\partial \cos\theta}{\partial y} = \pm 1$ distinguishes the two senses in which the magnetization rotates across the wall. The subscript $w$ labels quantities of the wall, to be distinguished from those of the meron below. Substituting \eqref{eq:Theta_eq} into \eqref{eq:TotalEnergy} gives the energy of a segment of length $l$ lying flat across the film, with the line tension

\begin{equation}
    {\sigma = 4h\sqrt{A K_z}}.
    \label{eq:tension}
\end{equation}

At this stage the energy does not depend on $\Phi$ and all in-plane orientations of the wall magnetization are degenerate.

At the second stage, the degeneracy is lifted by the
\textit{In-plane anisotropy}, a weaker easy axis $\hat{x}$ lying within the plane of the film, conditioned by $K_x \ll K_z$:

\begin{align}
U_{\mathrm{anisotropy,x}}
&= -K_x h \int d^2 r \, m_x^2
\notag\\
&= -K_x h \int d^2r \, \sin^2\theta\cos^2\Phi.
\end{align}
Integrating this across the wall profile \eqref{eq:Theta_eq}, with $\int dy\, \sin^2\theta = 2\lambda_w$, yields the segment energy
\begin{equation}
    U = 2hl\lambda_{w}\left(2K_z + K_x\sin^2\Phi \right) + \mathrm{const},
    \label{eq:dw_energy}
\end{equation}
minimized when $\Phi_0 = 0$ or $\pi$. For small deviations from a ground state, the energy density varies quadratically in $\delta\Phi = \Phi - \Phi_0$:
\begin{align}
U(\Phi)/l = U(\Phi_0)/l + \tfrac{\kappa}{2} (\Phi - \Phi_0)^2,
\end{align}
giving the stiffness for in-plane magnetization
\begin{equation}
    {\kappa = 4h\lambda_{w} K_x}.
    \label{eq:stiffness}
\end{equation}

We now allow the azimuthal angle to vary slowly along the wall such that $\Phi = \Phi(x)$. The gradient incurs the exchange cost from \eqref{eq:exchange}, adding $2Ah\lambda_w \int dx\, \Phi'^2$ to the energy. The resulting one-dimensional theory,

\begin{equation}
E[\Phi] = 2h\lambda_w \int dx \left[ A\, \Phi'^2 + K_x \sin^2\Phi \right] + \mathrm{const},
\label{eq:reduced_energy}
\end{equation}

presumes a separation of length scales

\begin{equation}
\lambda_w \ll \lambda_m \ll l.
\label{eq:scale_separation}
\end{equation}

The anisotropy hierarchy guarantees the first inequality, $\lambda_w/\lambda_m = \sqrt{K_x/K_z} \ll 1$, and the overall separation ensures that the wall profile \eqref{eq:Theta_eq} adjusts rigidly on the scale $\lambda_m$ over which $\Phi$ varies. 

Minimization of \eqref{eq:reduced_energy} yields

\begin{equation}
\frac{d^2\Phi}{dx^2}
=
\frac{K_x}{2A}\sin2\Phi.
\label{eq:phi_EL_eqn}
\end{equation}

A Bloch line is a region in which the azimuthal angle interpolates between the two ground states, $\Phi = 0$ and $\pi$. The interpolating solution is

\begin{align}
\cos\Phi(x) = \zeta_m \tanh \frac{x - X_m}{\lambda_m}, 
\quad \lambda_m = \sqrt{\frac{A}{K_x}},
\label{eq:phi_profile}
\end{align}

where $\lambda_m$ is the width of the meron and $X_m(t)$ is its position along the wall. $\zeta_m \equiv \tfrac{1}{2} \int_{-\infty}^{\infty} dx\, \frac{\partial \cos\Phi}{\partial x} = \pm 1$ is the corresponding $\mathbb{Z}_2$ topological charge. The meron's influence on the wall falls off exponentially at distances larger than $\lambda_m$, so the string parameters of a long segment, $l \gg \lambda_m$, are unaffected by its presence.

The in-plane anisotropy also renormalizes the wall width and tension at relative order $K_x/K_z$, and $\lambda_w \to \sqrt{A/(K_z - K_x\cos^2\Phi)}$ and $\sigma \to 4h\sqrt{A(K_z-K_x)}$. We retain these corrections where they matter quantitatively. The product $\sigma\kappa = 16h^2AK_x$, and with it the domain wall wave speed derived below, is unchanged.

For small deformations in the profile $Y(x,t)$, the energy varies as  

\begin{equation}
U = \sigma \int_0^l dx\, \sqrt{1+Y'^2}
\approx \sigma l+ \sigma\int_0^l dx\, \frac{1}{2}Y'^2,
\label{eq:tension_expansion}
\end{equation}

taking $Y' \ll 1.$ The Lagrangian describing a weakly perturbed domain wall is then comprised of a gyroscopic term and the above potential energy terms:

\begin{equation}
\mathcal{L} = 
g\dot{Y}\Phi
-
\frac{\sigma Y'^2}{2}
-
\frac{\kappa\left(\Phi-\Phi_0\right)^2}{2}.
\end{equation}

Minimization of the action with respect to the magnetization angle $\Phi$ yields

\begin{align}
\Phi = \Phi_0 + \frac{g}{\kappa}\dot{Y}.
\end{align}

Notice there is no coupling between $\Phi$ and $Y'$, so the elimination of $\Phi$ results in a perfectly reciprocal string; contrast this with the case of a wall with dipolar interactions, where such a coupling leads to nonreciprocal wave propagation \cite{Zhang:2018}.

Integrating out $\Phi$ yields a Lagrangian for the field $Y(x,t)$ only: 

\begin{align}
\mathcal{L} = \frac{\rho}{2}\dot{Y}^2 - \frac{\sigma}{2} (Y')^2,
\quad \rho = \frac{g^2}{\kappa}.
\label{eq:lagrangian}
\end{align}

In \cite{Zhang:2018}, it is established that this model is valid for wavelengths much larger than the domain wall width. Here, the wavelengths must also exceed the meron width. Retaining the gradient term of \eqref{eq:reduced_energy}, equivalently $\frac{\sigma\lambda_{w}^2}{2}(\Phi')^2$, produces the adjusted Lagrangian: 

\begin{equation}
\begin{split}
\mathcal{L} ={}& g\dot{Y}\Phi
-\frac{\sigma}{2}(Y')^2
-\frac{\kappa}{2}(\Phi - \Phi_0)^2 -\frac{\sigma\lambda_{w}^2}{2}(\Phi')^2 .
\end{split}
\end{equation}

This gives rise to the dispersion relation 

\begin{equation}
\omega^2
=c^2k^2\left(1+k^2\lambda_{m}^2\right),
\end{equation}

and the theory is recovered in the limit $k\lambda_{m}\ll1$.

\subsection{Configurations and Topological Charges}
\label{subsec:Topological Charge}
Three independent integers determine the Bloch line configuration. The two $\mathbb{Z}_2$ charges $\zeta_{\mathrm{m}}=\pm1$ in (\ref{eq:phi_profile}) and
$\zeta_{\mathrm{w}}=\pm1$ in (\ref{eq:Theta_eq}) have already appeared. The third integer, $\zeta_{\mathrm{c}}=\pm1$, is the polarity of the meron core (the orientation of spin along the $\hat{y}$-axis at the center of the line). There are therefore 8 configurations in total.

The Skyrmion charge of the meron is
\begin{equation}
Q
=
\frac{1}{4\pi}
\int dx\,dy\,
\mathbf{m}\cdot
\left(
\partial_x\mathbf{m}\times\partial_y\mathbf{m}
\right)
=
-\frac{1}{2}\zeta_{\mathrm{c}}\zeta_{\mathrm{m}}\zeta_{\mathrm{w}},
\label{eq:skyrmion_charge}
\end{equation}

a half-integer, as the meron is a half-skyrmion. The gyroscopic constant
follows from $Q$.

\begin{equation}
G \equiv G_{xy}=-G_{yx}
=
-4\pi Q \frac{M_s h}{\gamma}
=
\zeta_{\mathrm{c}}\zeta_{\mathrm{m}}\pi g,
\label{eq:gyro_constant}
\end{equation}

where $g = 2\zeta_w M_s h/\gamma = 2\zeta_w \mathcal{J}$, with $\mathcal{J}$ the angular-momentum density.

\subsection{Equations of Motion}
\label{subsec:EOM}

The Bloch line Lagrangian is built up from its components: two strings and a rigid meron. The meron is a point particle with coordinates $(x_m, y_m)$, and its Berry phase supplies a kinetic term of first order in the velocity, $\mathcal{L}_m = \mathbf{a} \cdot \dot{\mathbf{r}}_m$ \cite{PapanicolaouTomaras:1991}. Here $\mathbf{a}$ plays the role of a vector potential and we write it in the Landau gauge, $\mathbf{a}=(-Gy_m,\,0)$, although it could equally be replaced with $(0,\,Gx_m)$ or the rotationally symmetric $\frac{G}{2}(-y_m,\,x_m)$, which differ by a gauge transformation. Varying $\mathcal{L}_m$ produces the Lorentz force $\dot{\mathbf{r}}_m \times G\hat{z}$, revealing the meron to behave as a particle of unit charge in a background magnetic field $\nabla \times \mathbf{a} = G\hat{z}$, normal to the film. The strength and sign of $G$ are given by (\ref{eq:gyro_constant}). Note that the core polarity determines the sign of $G$, as it does for a magnetic vortex. 

The meron is constrained to slide along the domain wall. We implement the constraint with two Lagrange-multiplier terms, one per segment, as each string ends at the meron, and each endpoint must be pinned to the bead separately. The multipliers $\lambda_{1,2}$ are the transverse contact forces between the bead and the two strings. They need not be equal, and it is their imbalance that drives the meron. A single multiplier tying the segments to each other would enforce continuity of the wall but leave it decoupled from the bead. The full Lagrangian is then 

\begin{equation}
\begin{aligned}
\mathcal{L} ={}&
\int_{0}^{x_m(t)}\! dx\,
\left[
\frac{\rho_L}{2}\dot{y}_L^{\,2}
-\frac{\sigma_L}{2}(y_L')^2
\right]
\\
&+
\int_{x_m(t)}^{l}\! dx\,
\left[
\frac{\rho_R}{2}\dot{y}_R^{\,2}
-\frac{\sigma_R}{2}(y_R')^2
\right]\\
&- \sigma_Lx_m - \sigma_R(l-x_m)-G\dot{x}_m y_m
\\
&+
\lambda_1\!\left[y_m-y_L(x_m)\right]
+\lambda_2\!\left[y_m-y_R(x_m)\right],
\end{aligned}
\label{Lagrangian}
\end{equation}

 where $y_L(x,t)$ indicates the left domain wall segment and $y_R(x,t)$ the right. The potential term $U(x_m) = \sigma_Lx_m + \sigma_R(l-x_m)$ is an extension of (\ref{eq:tension_expansion}), arising from the difference in tension parameters, and discussed further in Sec. \ref{subsec:Bx}.

Minimization of the action (\ref{Lagrangian}) is subtle as the time-dependent bounds of integration generate kinetic terms in the equations of motion in addition to tension forces. These kinetic terms are derived in Appendix \ref{subsec:FullEOM}. Since they are higher order in the wave amplitude, guaranteed to be small by (\ref{eq:tension_expansion}), we retain them only when they simplify an expression without affecting the leading-order result as in (\ref{eq:xmKinks}).  Additionally, most perturbations leave $\sigma_L = \sigma_R$ and $\rho_L = \rho_R$, so these parameters will be taken as equal unless stated otherwise. The equations of motion are then

\begin{equation}
\begin{aligned}
G\dot x_m &= \sigma (y_R'- y_L')
\qquad\text{and}\qquad
G\dot y_m = \frac{\sigma}{2}(y_R'^2-y_L'^2).
\end{aligned}
\label{eq:eoms}
\end{equation}

 Away from the meron, the bulk of each domain-wall segment obeys the standard wave equation, 

\begin{equation}
    \rho \ddot y - \sigma y''  = 0, 
\end{equation}

with the open boundary conditions $y'(0,t) = y'(l,t) = 0$ at the film edge and $y(x_m) = y_m$ at the meron. 

\subsection{Meron as a Clamp}
\label{subsec:Clamp}

In the language of the wave equation, the meron is a common endpoint of the two segments-- a boundary that moves. In classic treatments of moving-boundary problems, the boundary's trajectory is prescribed and the string's response is computed \cite{Balazs:1961, Moore:1970, Dodonov:2001}. Here the boundary is itself dynamical. It moves in response to the tension of the strings through (\ref{eq:eoms}), so the string profile and the boundary trajectory must be determined together. We do so perturbatively, exploiting the built-in small parameter $y' \ll 1$. 

An exact solution for the profile of a string with a time-dependent support can be written in the form \cite{Law:1994}

\begin{equation}
    y(x,t) = \sum_n q_n(t)\, u_n(x, x_m(t)).
\end{equation}

 The small slope requirement suggests a small meron velocity compared to the string propagation speed, so we write $x_m(t) = x_0 + \delta x_m(t)$, where $x_0$ is the equilibrium position of the meron and $\delta x_m$ is its displacement, and expand the basis about $x_0$:

\begin{equation}
y(x,t) = \sum_{n} q_n(t)
\left[
u_n
+ \delta x_m\,\partial_{x_m} u_n
+ \frac{1}{2}\,\delta x_m^2\,\partial_{x_m}^2 u_n
+ \cdots
\right].
\label{eq:Wave_eqn_expansion}
\end{equation}

This expansion strictly works in the limit $k\,\delta x_m \ll 1$, and through a scaling analysis, we recover the adiabatic constraint on meron velocity $\dot{x}_m \ll c$. Additionally, we find the constraint on the wave amplitude $A \ll \frac{cG}{\sigma k}$. Thus, the meron speed will always be a fraction of the wave propagation speed under the influence of tension forces alone, and second order effects in amplitude have a minor impact on the meron trajectory. 

 To first order, the string profile is a superposition of plane waves. Writing $A$ for the incident amplitude, $B$ for the reflected, and $C$ for the transmitted, 

\[
y_L = A e^{i(kx - \omega t)} + B e^{-i(kx + \omega t)}, \quad
y_R = C e^{i(kx - \omega t)}.
\]

We find total reflection ($B/A = -1$, $C/A = 0$) such that the two sides are completely isolated, while the meron oscillates along the wall at the wave frequency,

\begin{align}
x_m(t) &= \frac{2 \sigma }{c G} A \, e^{-i \omega t}.
\label{eq:clamp_xm}
\end{align}

These results follow directly from the equations of motion (\ref{eq:eoms}). The longitudinal force on the meron is second order in the wave displacement, so to first order it vanishes, and with it the vertical response. Unable to follow the wave, the meron pins the wall at its own location. A point of the wall held fixed is a node, and a node reflects an incident wave completely, transmitting nothing. The transverse force, by contrast, is first order in the wave displacement, and it is this force that drives the horizontal oscillation (\ref{eq:clamp_xm}). The clamp is therefore not an assumption but a consequence of the gyroscopic dynamics. Put simply, moving the meron across the domain wall requires a force along it, and no such force exists at leading order. 

A second-order solution can be built from the first-order work if the displacement is expanded by orders, $x_m(t) = x_0 + \delta x_m^{(1)} + \delta x_m^{(2)}$, and (\ref{eq:Wave_eqn_expansion}) is updated accordingly. We will only state here that the second order boundary condition becomes

\begin{equation}
y_m(t)
=
-\frac{\sigma k^{2}A^{2}}{G}
\left[
t-\frac{\sin(2\omega t)}{2\omega}
\right].
\label{eq:ym_second_order}
\end{equation}

Notably, there is a small drift, linear in time, and a leakage to the other segment at twice the initial frequency. 

\section{External Perturbations}
\label{sec:perturbations}

We now drive the Bloch line with the standard experimental controls, namely magnetic fields directed along the easy and intermediate axes as well as spin-polarized currents, with two aims in mind. First, each drive probes a different element of the model: an out-of-plane field drives the wall as a whole and tests the gyroscopic coupling; an in-plane field exerts a force on the meron itself, splitting the string parameters and testing the junction conditions; currents couple to the meron directly. Second, each response reduces to a closed-form prediction, which Sec. \ref{sec:Mumax} confronts with micromagnetic simulations. We leave out a magnetic field along $\hat{y}$ as it does not meaningfully couple to the Bloch line except to bias the core polarity at the meron's inception.

\subsection{Magnetic field along easy axis}
\label{subsec:TransverseMagField}
A magnetic field parallel to the easy axis, $\mathbf{B} = (0, 0, B_z)$, breaks the energetic equivalence of the $\mathbf{m} = +\hat{z}$ and $-\hat{z}$ domains and thereby exerts the pressure (force per unit length)
$p_z = -2\zeta_w h M_s B_z$ on the domain wall. The pressure couples directly to the field $y$, adding a term $\mathcal{L_{\text{ext}}} = p_z y$ to the Lagrangian as described in \cite{Zhang:2018} and a constant drive to the wave equation $\rho \ddot y - \sigma y'' = p_z$. In the zero mode solution, this pressure is entirely balanced by the line tension such that $ - \sigma y''= p_z$. Of course, the domain wall also receives a kick at the instant the field is turned on, exciting oscillatory modes on top of this zero mode deformation. However, the contribution of the oscillatory components is small and averages to zero, and will not cause any mean displacement of the meron. The zero mode solutions for the left and right segments of the domain wall are then

\begin{equation}
y_L(x,t) = {\frac{p_z}{2\sigma} \left( x_m^2 - x^2 \right)},
\end{equation}

\begin{equation}
y_R(x,t) = {\frac{p_z}{2\sigma} \left( x_m^2 - x^2 \right)}
+ \frac{p_zl}{\sigma}(x - x_m).
\end{equation}

\begin{figure}[tbp]
    \centering
    \includegraphics[width=.9\columnwidth]{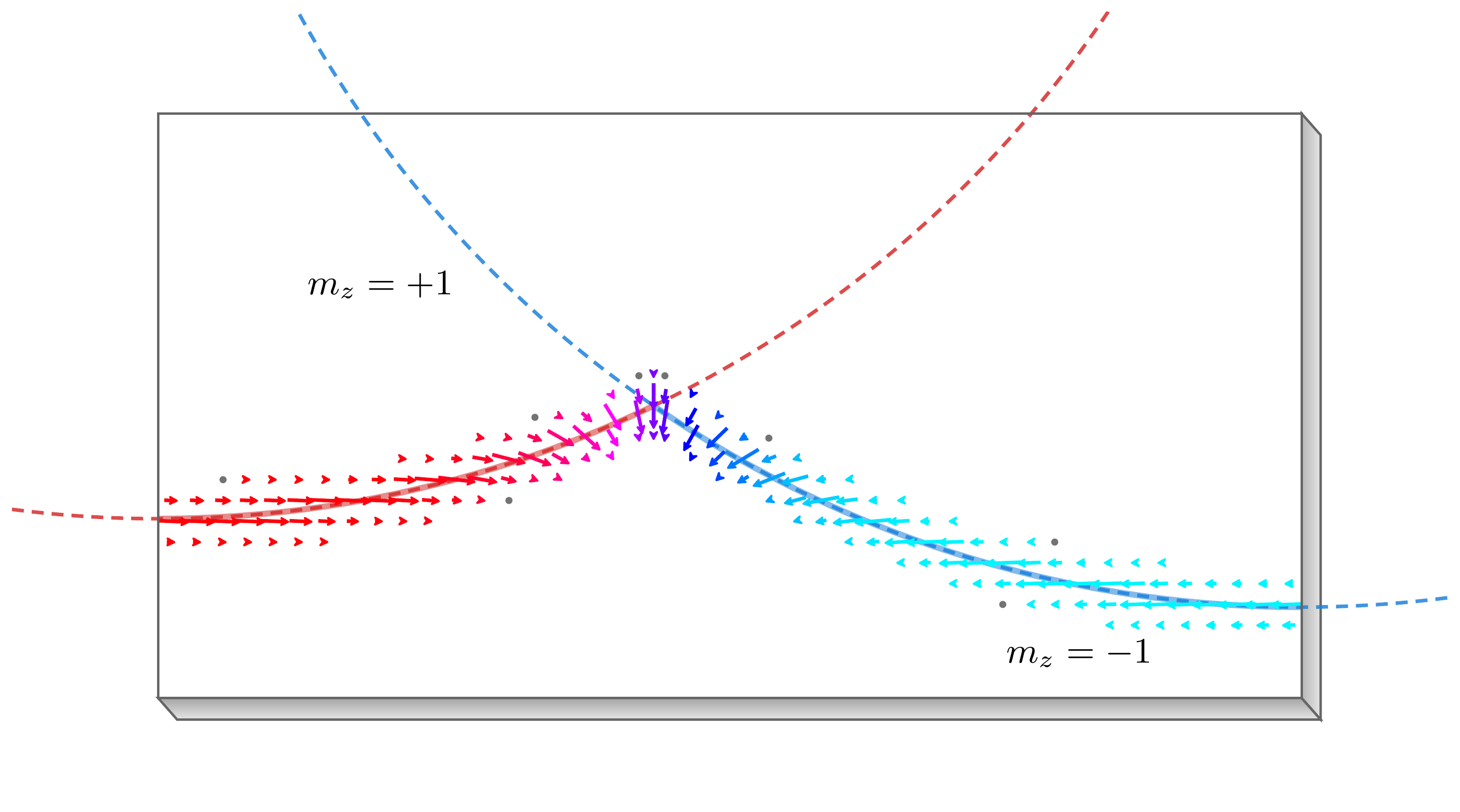}
    \caption{Meron on a domain wall at the intersection of two circles with exaggerated curvature. The circles are free to move up or down in order to keep the meron at the point of intersection as it travels across the film.}
    \label{fig:Bz_field_response}
\end{figure}

These solutions are parabolic approximations to a circular deformation, where each domain wall segment has been bent into a circle of radius $\frac{\sigma}{|p_z|}$, as illustrated by \textbf{Fig.} \ref{fig:Bz_field_response}. Each circle must conform to the zero slope boundary condition at the film edge, so there are only two degrees of freedom-- the y coordinates of the two circles' centers. These coordinates must adjust, and the radii must be large enough compared to the film length ($\frac{\sigma}{|p_z|} \gg l$), such that the meron is always located at an intersection of the circles. The problem of meron motion is thus re-framed as a geometric one. 

The velocity of the meron, averaged over time so that the oscillatory modes excited at turn-on drop out, is 

\begin{equation}
\left\langle \dot{x}_m \right\rangle
= \frac{\sigma}{G}\left(y'_R - y'_L\right)
= \frac{p_z l}{G}
= -\frac{\zeta_c\zeta_m\omega_L l}{\pi}
\end{equation}

The time to traverse the film will be a constant for any given magnetic field, regardless of film length. This is not a coincidence, but is a consequence of Larmor precession. 

Although $\Phi$ has been integrated out of the string dynamics, it remains a perfectly good observable, and its spatial average along the wall is set by the meron position alone. This average precesses at a rate given by the Larmor frequency $\omega_L = \gamma B_z$: 

\begin{equation}
\begin{aligned}
\left\langle \dot \Phi \right\rangle
    = \zeta_m \zeta_c\frac{\pi \dot{x}_m}{l}
     = - \omega_L.
\end{aligned}
\end{equation}

Composite domain walls often precess at approximately the Larmor frequency-- a vortex wall does, for example \cite{Clarke:2008}. However, this falls short of a theorem, which would require the conservation of $S_z$. Here the rate is exactly $\omega_L$, and is independent of $\zeta_c$. This independence comes from the fact that $\zeta_c$ dictates both the direction in which spins rotate across the meron, as well as the direction of $\dot x_m$, and these effects cancel. 

The y-coordinate velocity is calculated to be

\begin{equation}
    \langle \dot y_m \rangle = -\frac{p_z^{3}l^{2}}{G^{2}\sigma}\,t,
\end{equation}

where we set $x_m(0) = \frac{l}{2}$. By the time the meron reaches the film edge, at $t_f = \pi/(2 |\omega_L|)$, its net vertical displacement is $|y_m(t_f) - y_m(0)| = |p_z|l^2/(8\sigma)$. This is the order of the static sag of the wall itself, and small compared to the film size by the standing assumption $\sigma/|p_z| \gg l$. 

\begin{figure}[tbp]
    \centering
    \includegraphics[width=.7\columnwidth]{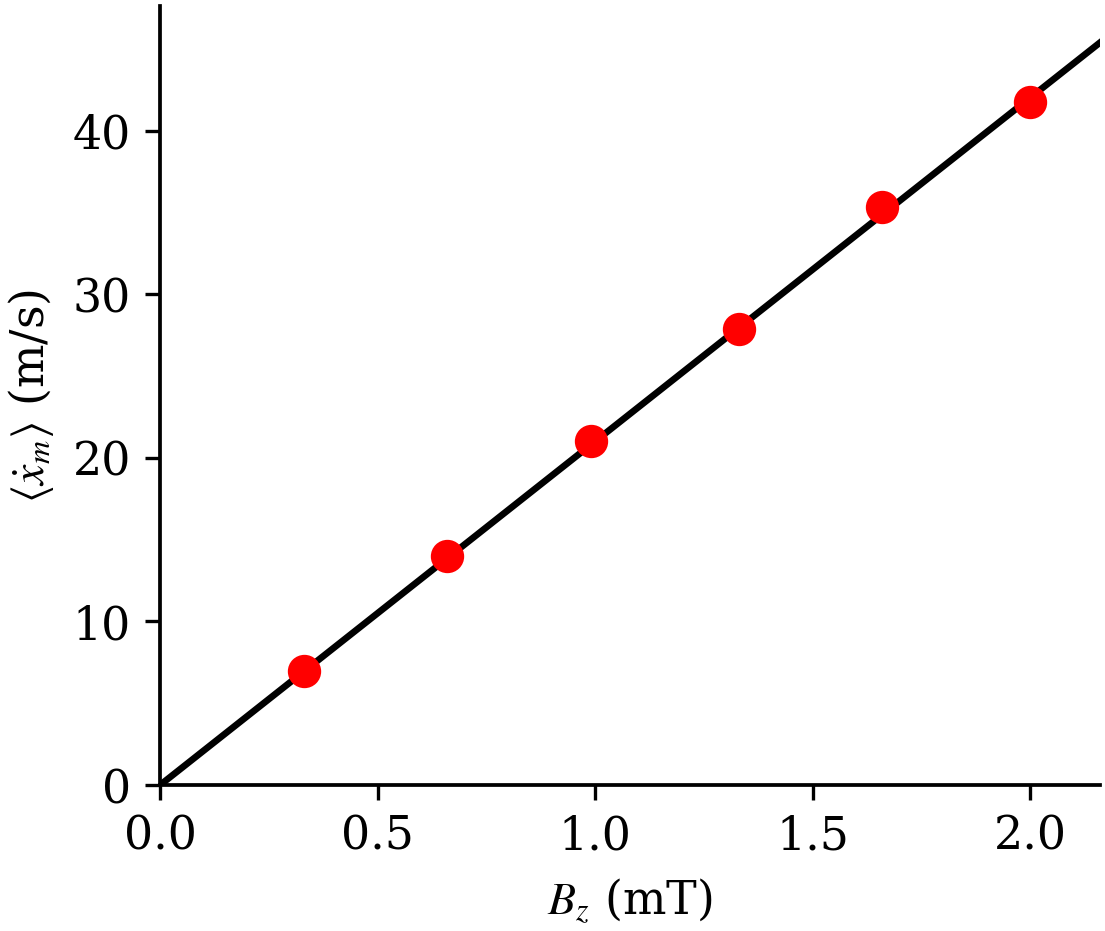}
    \caption{MuMax3 Simulations against theory for an out-of-plane magnetic field.}
    \label{fig:MumaxSimBz}
\end{figure}

In the strong-field, or infinite-length limit, an enormous tension force develops and the deformation becomes large enough that additional merons are nucleated. Comparison between theory and MuMax3 is illustrated in \textbf{Fig.} \ref{fig:MumaxSimBz}.

\subsection{Magnetic field along the intermediate axis}
\label{subsec:Bx}

\subsubsection{Domain Wall Parameters}

A magnetic field applied along the $\hat{x}$-axis, $\mathbf{B} = (B_x,0,0)$, breaks the energetic equivalence between +$\hat{x}$ and $-\hat{x}$ domains, and is by far the most illuminating perturbation to address. The magnetic field adds the Zeeman contribution $- M_sB_x \cos\Phi \sin\theta$ to the potential energy density, modifying both the energy per unit length and stiffness of the domain wall. Going forward, we define the small dimensionless parameter 

\begin{equation}
    \eta = \frac{M_s B_x}{2(K_z-K_x)}, 
\end{equation}

and extend our theory to $\mathcal{O}(\eta^2)$ in the meron velocities. Equation (\ref{eq:dw_energy}) becomes 

\begin{equation}
    U = 2h\lambda_{w}\left[
    2(K_z-K_x) + K_x\sin^2\Phi
    - \pi(K_z-K_x)\eta\cos\Phi
    \right].
    \label{eq:dw_energy_Bx}
\end{equation}

Expanding about $\Phi_0 = 0,\pi$ gives:
\begin{align}
    U
    &= 2h\lambda_{w}\Big[
    2(K_z-K_x) \mp \tfrac{\pi}{2} M_s B_x
    \nonumber\\
    &\qquad
    +\left(
    K_x \pm \tfrac{\pi}{4} M_s B_x
    \right)(\Phi-\Phi_0)^2
    \Big],
    \label{eq:Energydensity_zeeman}
\end{align}
where $(K_z-K_x)\eta = M_sB_x/2$ has been used to simplify, yielding

\begin{equation}
    \begin{gathered}
    \sigma_{\pm} = \sigma_0\left[1\mp\frac{\pi}{2}\eta\right],
    \qquad
    \kappa_{\pm} = \kappa_0\left[1\pm\frac{\pi K_z}{2K_x}\eta\right],
    \\
    c_{\pm} = \sqrt{\frac{\sigma_{\pm}\kappa_{\pm}}{g^2}}.
    \end{gathered}
    \label{eq:tension_stiffness_Bx}
\end{equation}

 $\sigma_0$ and $\kappa_0$ are the parameters in the absence of any field, stated in (\ref{eq:tension}) and (\ref{eq:stiffness}). The wall width (\ref{eq:Theta_eq}) is expanded out in the calculation of (\ref{eq:tension_stiffness_Bx}) and the $+$ subscript labels the parameters of the fast side. Evidently, the fast side is the one with reduced tension and increased stiffness. At higher orders in $\eta$, the spins cant such that $\theta \to \eta$ at $y \to +\infty$ and $\theta \to \pi - \eta$ at $y \to -\infty$, and the energy density (\ref{eq:Energydensity_zeeman}) must be re-derived.

\subsubsection{Meron Velocities}

With the splitting in speeds established, we now see that the overall problem is equivalent to that of two strings with different propagation speeds joined at a common, moving boundary:
\begin{equation}
    \rho \ddot{y} - \sigma y'' = 0,
    \qquad
    y'_L(0) = 0,
    \quad
    y'_R(l) = 0,
    \quad
    y(x_m) = y_m .
    \label{eq:wave}
\end{equation}

Without loss of generality we continue by taking $B_x > 0$ along with $\zeta_m = \zeta_c =  -1$. The difference in tension between the slow and fast sides provides the force

\begin{equation}
\begin{aligned}
    F_x
    &\equiv \sigma_R-\sigma_L
    = \sigma_--\sigma_+ \\
    &=  \pi \sigma_0 \eta
    = 2\pi h\lambda_w
    M_s B_x.
\end{aligned}
\end{equation}

The same driving force appears in the reduced Bloch-line equations of Zvezdin and Popkov \cite{ZvezdinPopkov:1986}. We emphasize that the Zeeman force only enters through the splitting of tension and stiffness parameters, and the sign is therefore physical rather than convention. In particular, tension will pull along the local tangent $\hat t = (\pm 1, \pm y')/\sqrt{1+y'^2}$, producing the force in $\hat{x}$:

\begin{equation}
    \frac{\sigma_R}{\sqrt{1+y_R'^2}} - \frac{\sigma_L}{\sqrt{1+y_L'^2}}
    \approx (\sigma_R - \sigma_L) + \tfrac{1}{2}\big(\sigma_L y_L'^2 - \sigma_R y_R'^2\big).
\end{equation}

The Zeeman force has thus manifested itself as $\sigma_R - \sigma_L$, the leading term. This asymmetry in tension drives the meron vertically at

\begin{equation}
\dot y_{m,d} = -\frac{F_x}{G} = -\zeta_w\omega_{L} \lambda_{w}, 
\end{equation}

with $\omega_L = \gamma B_x$. 

Because the meron is forever tied to the domain wall, as it moves vertically, it drags the rest of the wall along with it. Similar to a plucked guitar string, sharp discontinuities in the domain wall profile are generated. We refer to these discontinuities as kinks, visualized by the triangular markers in \textbf{Fig.} \ref{fig:Animated_DomainWallMotion}. Each kink now has a life of its own, one fast and one slow, propagating outwards towards the film edges. The remaining task in solving this perturbation has now become a matter of following these kinks and their evolution over time; the meron dynamics follow. 

Initially, the slopes of the left and right kinks are 

\begin{equation}
    y'_L = \frac{\dot y_{m,d}}{c_L + \dot x_m},
    \qquad
    y'_R = -\frac{\dot y_{m,d}}{c_R - \dot x_m}.
    \label{eq:initial_slopes}
\end{equation}

Upon plugging the above slopes into the full equation for $\dot x_m$ (\ref{eq:eom_x_corr}), the kinetic corrections cancel, and the stationary-boundary form survives: 

\begin{equation}
    \dot x_m=
    -\frac{ \dot y_{m,d}}{G}
    \left(
        \frac{\sigma_{L}}{c_{\mathrm{L}}}
        +
        \frac{\sigma_{R}}{c_{\mathrm{R}}}
    \right).
    \label{eq:xmKinks}
\end{equation}

\subsubsection{Reflection and Transmission of Kinks}

The shape of these kinks is preserved during free propagation. Even when they find the film edge, the kink's slope discontinuity will reflect with the same sign and magnitude. It is only during a collision with the meron that a kink may change its shape. Specifically, we expect the kinks to invert with a slightly altered magnitude since the meron acts as a moving, closed boundary, consistent with the clamping result of Sec. \ref{subsec:Clamp}.

To nail down a collision law for this altered magnitude we focus on the left side only. We write $y_L = f(x - ct) + g(x + ct)$, so $y'_L = f' + g'$ and $\dot y_L = c(g' - f')$. Exploiting the condition $\frac{dy_L(x_m,t)}{dt} = \dot y_m$ gives

\begin{equation}
    (\dot x_m - c)\, f' + (\dot x_m + c)\, g' = \dot y_m.
    \label{eq:kinbc}
\end{equation}

This relation, combined with the meron equations of motion (\ref{eq:eoms}),  simplifies to

\begin{equation}
    (y'_L - y'_R)^2 = \frac{2c_L G}{\sigma_L}(g' - f') - \frac{2 G}{\sigma_L}\dot y_{m,d}.
    \label{eq:collision_step}
\end{equation}

At a collision, the meron sees a jump in the left side slope such that $f' \xrightarrow{} f' + \Delta_I$. To ensure (\ref{eq:kinbc}) is still satisfied, $g$ must simultaneously jump from $g' \xrightarrow{}  g' + \Delta_R$, and the meron velocity changes from $\dot x_{m,1}$ before the collision to  $\dot x_{m,2}$ after. Subtracting (\ref{eq:collision_step}) before and after the collision gives 

\begin{equation}
    (\Delta_I + \Delta_R)(\dot x_{m,1} + \dot x_{m,2}) = 2c_L (\Delta_I - \Delta_R).
\end{equation}

We land at our collision law after rearranging $\Delta_R$ in terms of $\Delta_I$:

\begin{equation}
\Delta_R=\Delta_I\,\frac{c_L- \bar V}{c_L+ \bar V},
\label{eq:collision_law}
\end{equation}

 where $\bar V \equiv \tfrac12\big(\dot x_{m,1}+\dot x_{m,2}\big)$ is the average of the pre- and post-collision velocities.
 
 This result is similar to that in \cite{Balazs:1961} for a boundary moving at constant velocity, except our law uses a mean of the pre- and post-collision velocities. Importantly, this ensures that time-reversal symmetry is satisfied, expressed as 
 
\begin{equation}
    R(\bar{V})R(-\bar{V})=1,
    \qquad
    R(\bar{V})\equiv\frac{c-\bar{V}}{c+\bar{V}}.
\end{equation}

Expanding $\Delta_R = \Delta_I(1-2\bar V/c_0) + O(\eta^3)$, we see this symmetry is also preserved order by order in the expansion. Had this not been the case, energy would be spuriously pumped in, or siphoned from, the system at a collision. 

Finally, because the collision changes $\dot x_m$ and $\dot y_m$, the kinematic condition \eqref{eq:kinbc} on the opposite side is suddenly violated, so the meron must emit a small compensating kink into the far side. We denote this kink as a change in slope $\Delta_T$ and evaluate strictly to $\mathcal{O}(\eta^2)$:

\begin{equation}
    \Delta_T
    =
    \frac{G}{2\sigma_0c_0}
    \left(
    \dot x_{m,2}^2
    -
    \dot x_{m,1}^2
    \right).
    \label{eq:TransmissionLaw}
\end{equation}

\subsubsection{Implementation of Theory}

\begin{figure}[tbp]
    \centering
    \includegraphics[width=1\columnwidth]{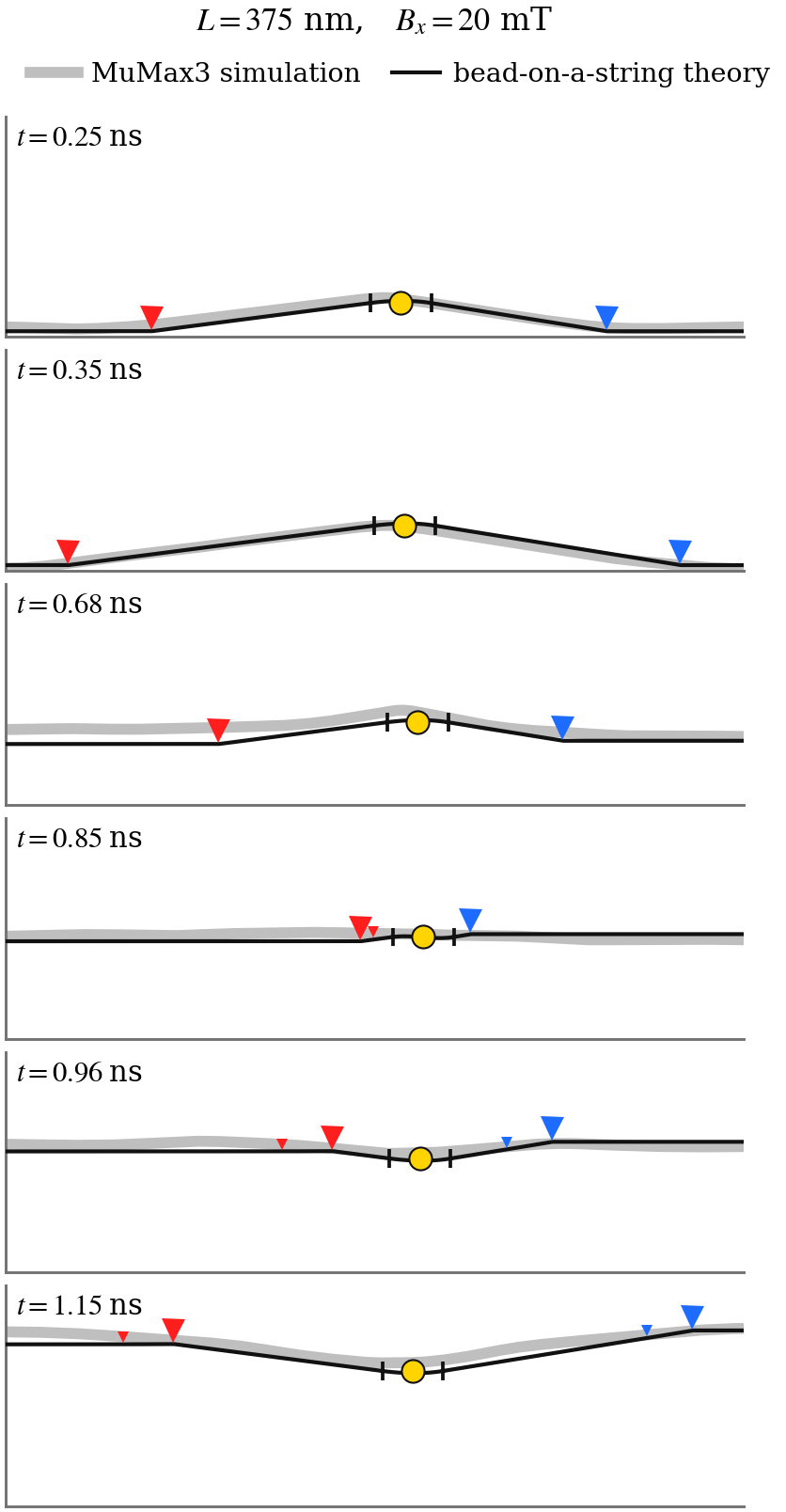}
    \caption{Piece-wise animation of the domain wall and meron according to our theory in a 20 mT magnetic field at a film length of 375 nm. The turning points used here are displaced from the core - a phenomenological correction to reconcile the meron's internal structure with our point-particle description. The red kink denotes the fast side and the blue represents the high tension slow side. At the moment the field is turned on, the meron will travel into the high-tension region, reducing the overall Zeeman energy. The smaller kinks appearing after $t=0.85$ ns are due to the secondary transmission upon a meron-kink collision. The thick gray outline is the extracted domain wall profile from MuMax3 simulations.}
    \label{fig:Animated_DomainWallMotion}
\end{figure}

Before turning to currents, we pause to test the collision framework assembled so far. We implement this theory in a Python script and compare the output to a set of MuMax3 trajectories. We find that for films of finite length, the idealization of the meron as a point-particle is imperfect. Kinks do not reach all the way to the core before reflecting. Instead, they turn at a point displaced from the core, and the collision is an extended scattering process beyond the point-particle model. We therefore treat the turning point as a phenomenological parameter, fitting it to $x_{turn} \approx x_m \pm(1.70 \pm 0.10)\lambda_m$ across external fields and film geometries. These displaced turning points can be seen in \textbf{Fig.} \ref{fig:Animated_DomainWallMotion}, represented as vertical bars around the meron. The natural length scale for this displacement, $ d =\ln{(1 + \sqrt{2})}\lambda_m$, is derived in Appendix \ref{sec:evanescence}, and is related to the fit value by a factor of two: $x_{turn} \approx x_m \pm 2d$. 

\begin{figure}[tbp]
    \centering
    \includegraphics[width=1\columnwidth]{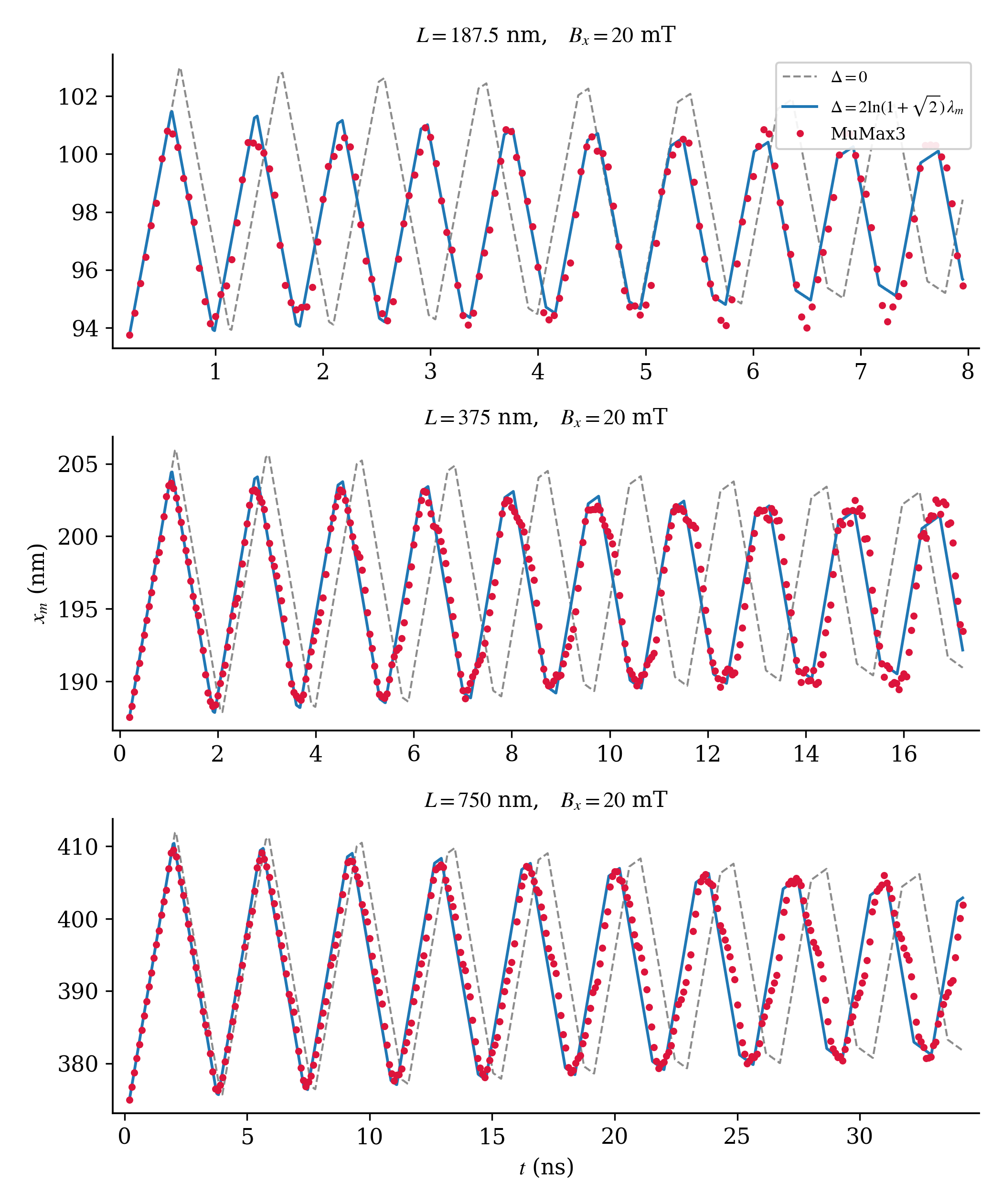}
    \caption{Meron motion in a 20 mT in-plane magnetic field for films of varying length. $\Delta = |x_{turn} - x_m|$ defines the turning point displacement from the core on each side.  The point-meron approximation converges onto the MuMax3 data with increasing film length, and the resulting trajectory resembles a triangular wave modulated by a slow envelope, most visible over the longest time window.}
    \label{fig:Bx_field_response}
\end{figure}

The trajectory level comparison with MuMax3 data at 20 mT is presented in \textbf{Fig.} \ref{fig:Bx_field_response}, where our point-particle theory converges onto the simulated trajectories as the film length increases. Animated snapshots of the domain wall profile, overlaid with the extracted MuMax3 profile for comparison, are shown in Fig. \ref{fig:Animated_DomainWallMotion}.

Beyond 20 mT, the increased domain wall velocity combined with decreased wall stiffness causes a gradual deviation between our theory and the MuMax3 simulations. In particular, $\Phi$ deviates enough from its ground state that it can no longer be integrated out, and a reliable theory must track both fields $\Phi(x,t)$ and $y(x,t)$. 

\subsection{Spin-Transfer Torque \texorpdfstring{$j_x$}{j\_x}}
\label{subsec:Jx}

Spin-transfer torques (STT) are the preferred mechanism for manipulating merons in proposed magnetic memory devices \cite{Humphrey:1985, S.S.P.Parkin:2008, Yang:2021, Nie:2025} and hence the most well-studied perturbation of Bloch lines. Per usual, we use the Thiele equation to capture the adiabatic torque \cite{Thiele:1973} and the dissipation function to capture the non-adiabatic component \cite{Thiaville2005, Zhang:2018}:

\begin{equation}
    (\mathbf{v} - \mathbf{u}) \times \mathbf{G}   + \beta \mathbf{D} \cdot \mathbf{u} + \mathbf{F} = 0.
    \label{eq:Thiele}
\end{equation}

For a current directed along $\hat{x}$, the meron's equations of motion become 

\begin{align}
    \dot{x}_m &= u_x + \frac{\sigma}{G}(y_R' - y_L') \\
    \dot{y}_m &= -\frac{\beta D_{xx}}{G} u_x + \frac{\sigma}{2G}(y_R'^2 - y_L'^2),
    \label{eq:STTeq}
\end{align}

where $u_x = \pi  \hbar  J_x / (q |G|)$ is the spin drift velocity. $J_x = Phj_x$ contains the spin polarization $P$, film thickness $h$, and current density $j_x$, $q = -e < 0$ is the electron charge (so $u_x < 0$ for a positive current: the texture drifts along the electron flow), and $D_{xx} = 2|g| \lambda_w/\lambda_m$ is the dissipation coefficient.

Setting tension aside for a moment, we see that the meron is driven at  

\begin{equation}
\dot x_{m,d} = \frac{ \pi  \hbar J_x}{q |G|}
\label{eq:STTeq_x}
\end{equation}
and
\begin{equation}
\dot y_{m,d} = -\frac{ 2 \beta \hbar J_x \lambda_w}{q G \lambda_m}.
\label{eq:STTeq_y}
\end{equation}

\begin{figure}[tbp]
    \centering
    \includegraphics[width=1\columnwidth]{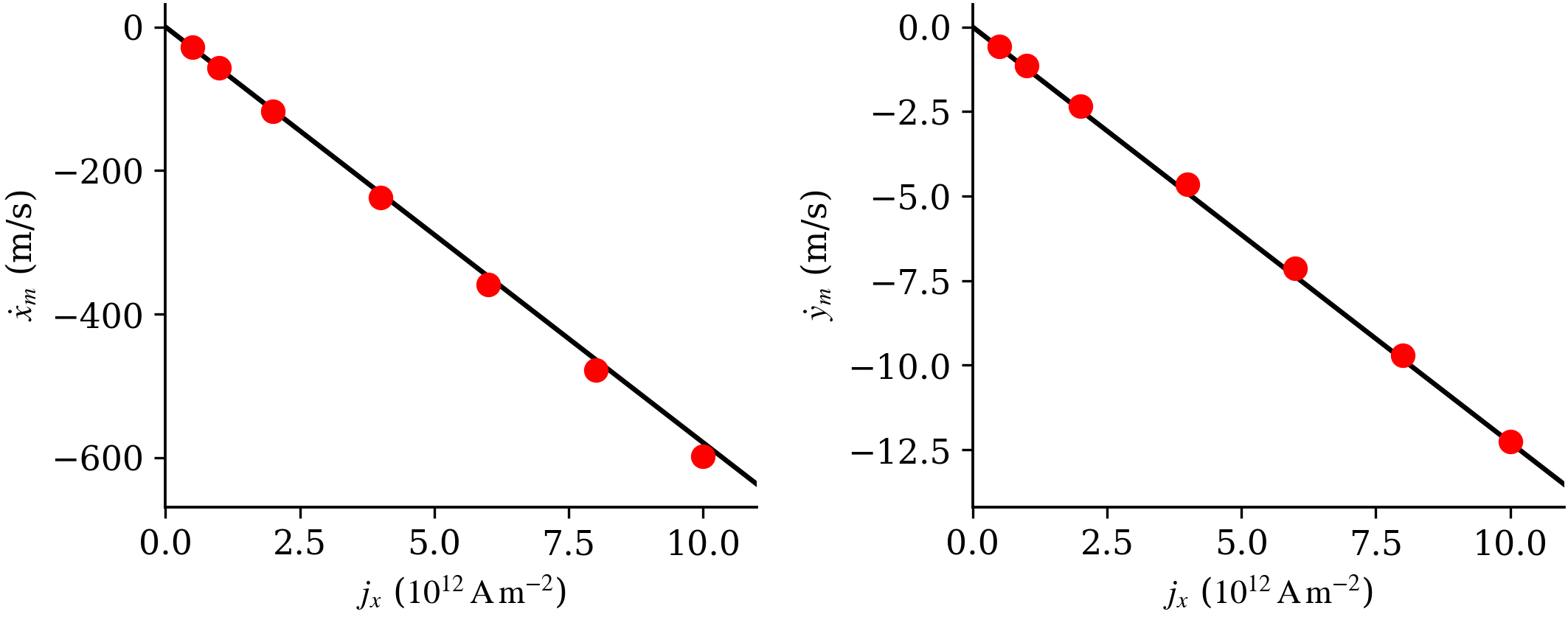}
    \caption{Meron velocity due to a spin current along the $\hat{x}$ axis.}
    \label{fig:STTx_field_response}
\end{figure}

 In the adiabatic limit of $\beta = 0$, the meron will ride the spin current, traveling through the domain wall at constant velocity, regardless of the sign on $G$, while the bulk domain wall is unaffected. Since there are no deformations in the domain wall, the meron can far exceed $c$ without any nucleation occurring (this robust behavior is in agreement with \cite{Herranen:2015, Herranen:2017, Yang:2021}) and is verified in \textbf{Fig.} \ref{fig:STTx_field_response} against MuMax3 simulations.  Thus, sending a current along the bulk domain wall will move the meron in the fastest manner possible. We note (\ref{eq:STTeq_x}) and (\ref{eq:STTeq_y}) are similarly recovered in the charge-as-coordinate construction of \cite{Tchernyshyov:2023}, most suitable if the reciprocal spin-motive force produced by the fast Bloch line is of interest.

When $\beta$ is nonzero, the meron's motion along $y$, a gyrotropic response dependent on $G$, will emit kinks into the bulk domain wall as in the case of an applied magnetic field along the $\hat{x}$-axis. However, the tension forces exerted back onto the meron are several orders of magnitude less than the primary velocity drive and are not worth stating (the handling of kinks is also a more challenging problem than that discussed in Sec. \ref{subsec:Bx} since the meron velocity can be supersonic).

\subsection{Spin-Transfer Torque \texorpdfstring{$j_y$}{j\_y}}

We may also apply a current in the $\pm \hat{y}$ direction, a perturbation which couples directly to both the domain wall and meron. Following \cite{Zhang:2018, Thiaville2005}, this STT exerts a pressure and a torque on the domain wall, giving rise to the external contribution $\mathcal{L}_{\text{ext}} = \int dx\left(\tau \Phi + p_{\beta}y\right)$, with $p_{\beta} = \frac{\hbar \beta J_y}{q\lambda_w}$, $\tau = -g u_y$, and $u_y = \frac{ \pi  \hbar J_y}{q |G|} $. The meron's response is determined by the Thiele equation and dissipation function of (\ref{eq:Thiele}). The domain wall profile is then governed by the below system:  

\begin{equation}
\begin{aligned}
    \rho\ddot{y} - \sigma y''
        &= p_{\beta},
        \qquad y(x_m,t) = y_m(t) , \\
    \partial_x y_L(0,t)
        &= \partial_x y_R(l,t) = 0,
        \qquad \dot y(x,0) = -\frac{\tau}{g} = u_y.
\end{aligned}
\end{equation}

The meron velocities are 

\begin{equation}
\left\langle \dot{x}_m \right\rangle = \frac{p_{\beta} l}{G} = \frac{ \beta \hbar J_y l}{ q G \lambda_w} 
\label{eq:meronJy_xm}
\end{equation}
 and 
\begin{equation}
\left\langle \dot y_m \right\rangle = \frac{ \pi  \hbar J_y}{q |G|}.
\end{equation}

The complete response of the Bloch line to this current now has a simple understanding: When $\beta = 0$, the meron and domain wall undergo uniform translation, moving together. When $\beta \neq 0$, the domain wall will deform into a parabola exactly as it did for an out-of-plane magnetic field, with $p_z \rightarrow p_{\beta}$. The dissipation term of (\ref{eq:Thiele}) is intended to be an equation local to the meron, where $D_{yy}$ is integrated across the length of the meron rather than the domain wall as a whole. With this interpretation, the contribution due to dissipation is negligible compared to tension at order $\lambda_m/l$. If $D_{yy}$ instead accounts for the entire domain wall length, then $D_{yy} = |g| l/ \lambda_w$, and $\frac{\beta D_{yy}}{G} u_y = \frac{p_{\beta} l}{G}$ naturally encodes the tension term. 

\section{Micromagnetic Simulations}
\label{sec:Mumax}

We use the MuMax3 micromagnetic solver \cite{MuMax} to simulate the dynamics of a Bloch line. Our parameters are representative of those reported for CoFeB thin films, with comparable values of $M_s$, $A$, and $\gamma$ \cite{CoFeBParameters}. The material parameters are: 
$M_s = 1 \times 10^6~\mathrm{A/m}$, 
$A = 1 \times 10^{-11}~\mathrm{J/m}$,  
$K_z = 1.3 \times 10^6~\mathrm{J/m^3}$, 
$K_x = 1.3 \times 10^5~\mathrm{J/m^3}$, 
$\gamma = -176 \times 10^9~\mathrm{rad/(s \cdot T)}$ (MuMax3 takes the magnitude $|\gamma|$ as input, with the electron's sense of precession built into its torque expression), P = 1, $\beta = 0.1$, and $\alpha = 0$. We set $\zeta_w = 1$ and  $\zeta_m =\zeta_c = -1$ so the product $\zeta_c\zeta_m \zeta_w = +1$ is positive. 

The domain wall coordinates are extracted by fitting the z-coordinate of magnetization, $\cos{\theta}$, to a $\tanh{\frac{y-Y}{\lambda_w}}$ profile along every vertical strip of the film, and extracting the location of the domain wall $Y$. The meron coordinates are similarly found by fitting a $\sech{\frac{x-X_m}{\lambda_m}}$ profile to the y-coordinate of magnetization across the extracted domain wall lattice sites. The velocity is determined by fitting a line to the meron's x and y coordinates over time. 

 For the simulations of the out-of-plane magnetic field and spin transfer torque, the geometry of our film is held fixed at $375~\mathrm{nm} \times 50~\mathrm{nm} \times 0.625 ~\mathrm{nm}$ (x, y, z), and the meron is initialized in the center. The duration of the simulation depends on the strength of the applied perturbation, and we track the trajectory from just after the perturbation is turned on to just before the meron exits the film. For the in-plane magnetic field, we vary the geometry of the film, in both the $\hat{x}$ and $\hat{y}$ directions. Discretization errors were observable and had a real impact on our results. To reduce them, we use the small cell size of $0.625$ nm/cell. We relax the simulation with damping, $\alpha = 0.5$, before turning the perturbation on and setting $\alpha = 0$. 

\section{Discussion}

In this paper, we have established that the meron behaves as a gyroscopic bead on a string, a point-like boundary that moves under the tension of the domain wall it inhabits. Its central property is the clamp effect: at low amplitudes the meron completely reflects the wall's spin waves, dynamically isolating the segments on its left and right. It further pins the domain wall in place, modifying the allowed modes of each segment. The responses to the standard drives are unified through this clamped junction description, and the solutions to each perturbation boil down to solving the wave equation with a pressure term, driven boundary, or both. 

The perturbation which puts our model most to the test is the in-plane magnetic field. We find, counter-intuitively, that the Zeeman force sneaks into the equations of motion as the leading term in tension. As a result, the sign of the force is not merely a convention but carries a physical meaning-- energy is only conserved when the appropriate sign is used, shown in the Appendix \ref{subsec:econserve}. The in-plane field also generates sharp discontinuities in the domain wall profile, kinks, which bounce between the film edge and meron. We derive a Doppler-like law for how these kinks transform under a collision with the moving meron, analogous to laws found elsewhere in the study of waves and recoiling boundaries \cite{Balazs:1961, Barnett:2010}. 

Comparisons to micromagnetic simulations reveal where the finite width of the meron is important, leading to the inclusion of a phenomenological parameter in our theory. This parameter stands in for a full wave-scattering treatment at the meron core, which we leave to a future work. As the film length increases, this parameter becomes less and less necessary, and the predictions of our theory are recovered.  

A natural next step is to move toward a more realistic description of the Bloch line, where long-range dipolar interactions and damping must be considered. The inclusion of dipolar interactions will lead to nonreciprocal wave propagation, dispersion, and edge repulsion as previously discussed in \cite{Zhang:2018, Makhfudz:2012}. Additionally, the boundary conditions at both the meron and film edges will satisfy a new, undetermined relation. Gilbert damping adds Rayleigh dissipation terms to (\ref{eq:eoms}), leading to a transmission coefficient of order $\alpha$ rather than zero. Finally, extending the model to multiple merons in order to study the interactions between neighboring defects on a common domain wall is of interest, and follows from \cite{MantonSutcliffe:2004Kinks, Nikiforov1986BlochLines, MalozemoffSlonczewski1979}. Such an extension would also open the door to meron nucleation at the domain wall's velocity limit and its relation to the dynamic multiplication of Bloch lines \cite{ZvezdinPopkov:1986}.

\section{Acknowledgments}

J.H. acknowledges the support of a Provost's Undergraduate Research Award (PURA) at Johns Hopkins University, awarded in Fall 2025. The authors used AI assistants (Anthropic's Claude and OpenAI's ChatGPT) for the numerical implementation of the meron-kink collision model, independent verification of analytical results, and figure preparation. All results were derived and verified by the authors.

\section{Appendix}

\subsection{Complete Equations of Motion}
\label{subsec:FullEOM}

The full equations of motion for the meron are 
\begin{align}
G\dot x_m &= \sigma_R y_R'-\sigma_L y_L'
\nonumber\\
&\quad +\dot x_m\big[\rho_R\dot y_R-\rho_L\dot y_L\big],
\label{eq:eom_x_corr}\\[4pt]
G\dot y_m &= -F_x + \tfrac12\big[\sigma_R (y_R')^2-\sigma_L (y_L')^2\big]
\nonumber\\
&\quad +\tfrac12\big[\rho_R\dot y_R^2-\rho_L\dot y_L^2\big]
\nonumber\\
&\quad +\dot x_m\big[\rho_R\dot y_R y_R' -\rho_L\dot y_L y_L'\big].
\label{eq:eom_y_corr}
\end{align}

 The terms associated with velocity become familiar if we integrate the x and y momenta over a string. The transverse momentum of the left segment is 

\begin{equation}
    P_y = \int_{0}^{x_m(t)} dx \rho_L\dot y_L.
\end{equation}

Differentiating with respect to time explains the origin of $\rho \dot y \dot x_m$: 

\begin{equation}
\begin{split}
    \frac{dP_y}{dt}
    &= \int_{0}^{x_m(t)} dx\, \rho_L \ddot y_L
     + \rho_L \dot y_L\dot x_m \\
    &= \sigma_L \, y_L'\big|_{0}^{x_m(t)}
     + \rho_L \dot y_L\dot x_m .
\end{split}
\end{equation}

The terms evaluated at the meron play a role in the equations of motion while the terms at the film edge are associated with a momentum flux at that boundary.

The momentum along x is given by 

\begin{equation}
    P_x = \int_{0}^{x_m(t)} dx \rho_L\dot y_L y_L' = \int_{0}^{x_m(t)} dx T^{tx},
\end{equation}

where $T^{tx}$ refers to the momentum density along x obtained from the energy-momentum tensor for a string. Using conservation of momentum,
\begin{equation}
\partial_t T^{tx}
+
\partial_x T^{xx}
=
0 ,
\end{equation}

we may evaluate the time derivative $\dot P_x$  as 

\begin{align}
\frac{dP_x}{dt}
&=
\int_0^{x_m(t)} dx\,
\partial_t T^{tx}
+
\dot x_m T^{tx}\big|_{x_m}
\nonumber\\
&=
\left[
\frac{\rho_L}{2}\dot y_L^2
+
\frac{\sigma_L}{2}(y_L')^2
\right]_{0}^{x_m}
+
\rho_L \dot x_m \dot y_L y_L' .
\end{align}

Examining $\dot P_y$ and $\dot P_x$ in combination with the open boundary condition, it is clear that momentum may enter the string either through transverse movement at the film boundary or through the meron's movement. The kinetic terms then enter as a consequence of momentum transfer between the left and right sides. 

Lastly, if the string parameters between the left and right sides are equal, a nice simplification occurs: 

\begin{align}
G\dot{x}_m
&=
\sigma \left(y_R'-y_L'\right)
\left[
1-\left(\frac{\dot{x}_m}{c}\right)^2
\right],
\label{eq:xcollapse}\\
G\dot{y}_m
&=
\frac{\sigma}{2}\left(y_R'^2-y_L'^2\right)
\left[
1-\left(\frac{\dot{x}_m}{c}\right)^2
\right].
\label{eq:ycollapse}
\end{align}

\subsection{Conservation of Energy}
\label{subsec:econserve}

We perform an independent consistency check on both the sign of $F_x$ and the complete equations of motion by verifying that these dynamics conserve energy. The conserved energy is

\begin{equation}
    E = \sum_{\text{sides}} \int \Big[\tfrac{\sigma}{2}y'^2 + \tfrac{\rho}{2}\dot y^2\Big] dx
    + \sigma_L x_m + \sigma_R (l - x_m) .
\end{equation}

The only site where $dE/dt$ may be nonzero is at the meron boundary: 

\begin{equation}
  \frac{dE}{dt} \;=\; \dot x_m\left(\sigma_L-\sigma_R\right)
    + \dot x_m\left(\varepsilon_L - \varepsilon_R\right)
    + \sigma_L y'_L \dot y_L - \sigma_R y'_R \dot y_R,
  \label{eq:power}
\end{equation}

where we write the energy density evaluated at the meron as $\varepsilon = \tfrac{\sigma}{2} y'^{\,2}
+ \tfrac{\rho}{2}\dot y^{\,2}$. Plugging in the full equations of motion (\ref{eq:eom_x_corr}) and (\ref{eq:eom_y_corr}), we are left with 

\begin{equation}
\begin{aligned}
\frac{dE}{dt}
&= \dot{x}_m \Big[
\rho_R \dot{y}_R
\left(\dot{y}_R + \dot{x}_m y'_R - \dot{y}_m\right) \\
&\qquad
- \rho_L \dot{y}_L
\left(\dot{y}_L + \dot{x}_m y'_L - \dot{y}_m\right)
\Big] = 0.
\label{eq:conserved}
\end{aligned}
\end{equation}

This vanishes identically by the chain rule applied to the constraint $y(x_m(t),t) = y_m(t)$, which gives $\partial_t y(x_m,t) = \dot{y}_m - \dot{x}_m\,y'$ on each side of the meron; all energy is accounted for. 

\subsection{Meron as an Extended Boundary}
\label{sec:evanescence}

We now examine the internal structure of the meron to understand why kinks reflect before reaching the core. In the vicinity of the core, the spins rotate away from $0$ and $\pi$, and the propagation speed becomes 

\begin{equation}
    c^2(\Phi) = \frac{\sigma(\Phi)\kappa(\Phi)}{g^2} \approx c_0^2 \cos 2\Phi.
\end{equation}

Solving in terms of x, we have 

\begin{equation}
    c^2(x) = c_0^2\left(1 - 2\sech^2\frac{x - x_m}{\lambda_m}\right). 
\end{equation}

At the positions

\begin{equation}
x = x_m \pm \ln(1 + \sqrt{2})\lambda_m,
\label{eq:turningpoint}
\end{equation}

or equivalently $\Phi =  \pi/4$ and $3\pi/4$, the D{\"o}ring mass diverges, and the domain wall becomes infinitely heavy such that $c^2(x)=0$. For $\pi/4 < \Phi < 3\pi/4$, we have $c^2(\Phi)<0$, and the meron acts as an evanescent barrier. A kink, being a sharp broadband excitation, will carry momentum into the meron, climb up the potential-energy barrier, and reflect once propagation ceases. Each turning point is consequently displaced from the meron core by the distance $d=\ln(1+\sqrt{2})\lambda_m$. 

This derivation establishes a natural length scale for the displacement, but it cannot be directly inserted into the point-particle theory we have established. The equations of motion, the kink slopes, and the collision laws all evaluate string quantities at $x_m$, implicitly assuming the wave field extends all the way to the core. However, the presence of the evanescent barrier contradicts this since waves cannot propagate within a distance $d$ of the core. A self-consistent treatment would need to reformulate the way tension is applied to the meron and the results that follow, which we leave to a future work.

We therefore treat the turning point as a phenomenological parameter, fitted to MuMax3 data. The fitted value $x_{turn} \approx x_m \pm 2\ln(1+\sqrt{2})\lambda_m$ is approximately twice the derived turning point, indicating that additional physics beyond the evanescent barrier picture is involved in the meron-kink scattering process.

\bibliography{main}

\end{document}